\documentclass[11pt]{article}
\ifdefined\reviewversion
  \usepackage[review]{acl}
\else
  \usepackage[preprint]{acl}
\fi
\usepackage{times,latexsym}
\usepackage[T1]{fontenc}
\usepackage[utf8]{inputenc}
\usepackage{microtype,courier}
\usepackage{graphicx,booktabs,amsmath,amssymb,array,tabularx,makecell}
\newcolumntype{Y}{>{\raggedright\arraybackslash}X}

\newcommand{\Dadv}{\Delta P_{\mathrm{adv}}}
\newcommand{\Dsafe}{\Delta P_{\mathrm{safe}}}
\newcommand{\ind}{\mathbf{1}}
\newcommand{\NA}{\textnormal{n/a}}
\newcommand{\NR}{\textnormal{not reported}}

\title{Decision Hijacking: Prompt Injection Attacks on Jev's Typed Probabilistic Decisions}
\author{Tiantong Wu \\
  Nanyang Technological University \\ Singapore \\
  \texttt{tiantong.wu@ntu.edu.sg}
  \And
  Wei Yang Bryan Lim \\
  Nanyang Technological University \\ Singapore \\
  \texttt{bryan.limwy@ntu.edu.sg}}

\begin{document}
\maketitle

\begin{abstract}
Most studies of prompt injection focus on generative agents, leaving their effects on models with schema-defined outputs unclear. We examine these effects in Jev, a non-generative decision model, using 510 reconstructed InjecAgent cases. Malicious content shifts action probabilities but rarely causes Jev to select the attacker's target. Override markers reduce this influence, while claims of contextual relatedness have small effects. Adaptive attacks using score feedback double the mean highest attacker-target probability found during optimization, while success on fresh validation calls rises from 1.8\% to 3.5\%. Exploratory analysis links these successes to small initial decision margins or greater attacker control over the observation. Together, these findings show that schema-defined outputs change but do not eliminate prompt-injection risk, highlighting the need to evaluate how untrusted content influences choices within the allowed action set.
\end{abstract}

\section{Introduction}
\label{sec:introduction}
Large language model (LLM) agents combine language understanding with tool use to retrieve information and act on their environment~\citep{yao2023react,schick2023toolformer}. These capabilities require models to process external content, including webpages, documents, emails, and tool responses. Such content may contain instructions that conflict with the user's task. Indirect prompt injection exploits this exposure by placing malicious instructions in data that the agent later reads~\citep{greshake2023indirect,yi2025bipia}. The resulting failures can include selecting an attacker-target tool, abandoning the user's request, or disclosing private information~\citep{zhan2024injecagent,debenedetti2024agentdojo}. Understanding when external content can redirect an automated decision is therefore important for systems that act on behalf of users.

Existing research has established this risk in generative LLM systems. Benchmarks evaluate attacks through generated responses and agent actions~\citep{yi2025bipia,zhan2024injecagent,debenedetti2024agentdojo}, while tool-selection attacks show that malicious descriptions can influence which available tool a generative agent chooses~\citep{shi2025toolhijacker,sneh2025tooltweak}. Defenses seek to preserve the distinction between trusted instructions and untrusted data through structured inputs or model training~\citep{chen2025struq,chen2025secalign}. These findings do not establish how the same attacks behave when the decision model does not generate text and can only select from a predefined structured output space.

TypeSafe AI's Jev provides a concrete setting for this question. Its \emph{choice} interface maps a state and a predefined set of options to a typed decision and reported probabilities, without generating free-form text~\citep{typesafe2026jev,typesafe2026api}. The interface restricts selection to declared actions, but an available action can still conflict with the user's task. We call the remaining risk \textbf{decision hijacking}, in which attacker-controlled content shifts the decision distribution toward an attacker-target action. For state $s$, fixed action set $\mathcal A$, safe-action set $A_{\mathrm{safe}}\subseteq\mathcal A$, and attacker-target set $A_{\mathrm{adv}}\subseteq\mathcal A$, an attack replaces $s$ with $s'$ and changes
\begin{equation}
 p(a\mid s,\mathcal A) \longrightarrow p(a\mid s',\mathcal A).
 \label{eq:attack}
\end{equation}
The user's task and available actions remain unchanged. We examine both attacker-target selection and probability shifts, since the latter may provide useful optimization feedback even when the selected action does not change.

To study this gap, we reconstruct the 510 direct-harm combinations in InjecAgent as typed decisions immediately after exposure to malicious content. The reconstruction preserves the original user goals, attacker goals, injection channels, and attack text, while replacing free-form generation with a closed set of choices. Four experiments cover 54060 calls in the main study. DH-1 tests whether original attacks transfer to Jev. DH-2 separates the effects of injection-style markers. DH-3 tests assertions of contextual relatedness against a registered smallest effect size of interest (SESOI). DH-4 uses a deterministic grammar-based optimizer with access to the reported probability vector. Following the need for adaptive evaluation established in prior work~\citep{zhan2025adaptive,hofer2026automated}, this final experiment tests whether score feedback can turn small probability shifts into validated attacker-target selections. We evaluate each selected attack with five fresh calls, separately from the calls used during optimization. Validated success requires at least three of these calls to select the attacker target.

The results distinguish changes in probability from changes in action selection. Original attack content increases mean attacker-target probability by 0.043, but selects the target in only 1.8\% of cases. Adding the source override prefix weakens the attack. In DH-2, adding ``ignore all previous instructions'' lowers attacker-target probability by 0.032 relative to the plain request. In DH-3, the stronger relatedness assertion produces small positive contrasts that remain below the registered SESOI of 0.02. In DH-4, 24 additional queries double the mean best-so-far attacker probability from 0.043 to 0.084, while validated success rises from 1.8\% to 3.5\%. Familiar injection markers and adaptive score access, therefore, affect Jev's decision distribution without producing widespread validated hijacking in this benchmark.

Exploratory analysis identifies two patterns associated with successful adaptive attacks. Some decisions have a small initial gap between safe-action and attacker-target probability. In other cases, the attacker controls most of the observations and can induce a larger change in probability. Among cases with a large initial margin and malicious content embedded in a larger observation, only 1 of 468 is successfully hijacked. These associations describe the observed run and do not establish universal causal rules.

The study contributes a threat model and benchmark reconstruction for a non-generative typed decision layer, controlled evidence separating probability shifts from validated action hijacking, and an exploratory analysis of the conditions associated with success. The findings show that restricting outputs to declared actions does not by itself establish resistance to prompt injection. Evaluation should also measure changes in probability, decision margins, observation structure, and the effect of exposing scores to an adaptive attacker. Our claim concerns the transfer of attack content to the tested Jev interface, rather than an overall security ranking of generative and non-generative models.

\section{Background and Related Work}
\label{sec:related}

\paragraph{Tool use and indirect prompt injection.}
ReAct combines generated reasoning with actions that obtain information from the external environment~\citep{yao2023react}. Toolformer trains a language model to decide when to call APIs and how to incorporate their results~\citep{schick2023toolformer}. These approaches illustrate how external content enters a model's decision process. \citet{greshake2023indirect} shows that malicious instructions in retrieved content can redirect LLM-integrated applications. BIPIA evaluates indirect prompt injection across tasks involving external content and studies defenses that help models distinguish data from instructions~\citep{yi2025bipia}. Our study retains this threat from untrusted input while examining a non-generative decision interface.

\paragraph{Benchmarks for agent security.}
InjecAgent provides 1054 cases covering 17 user tools and 62 attacker tools, with direct-harm and data-stealing objectives~\citep{zhan2024injecagent}. Its enhanced condition adds an override phrase to the original attacker instruction. AgentDojo evaluates attacks and defenses in stateful tool environments with task and security checks~\citep{debenedetti2024agentdojo}. We reconstruct InjecAgent's direct-harm cases because they provide paired user tasks, attacker goals, tool-output templates, and attacker-target tools. Our endpoint is selection at the first security-critical decision after exposure to the injected content. It does not measure the completion of an end-to-end harmful workflow.

\paragraph{Attacks on tool selection.}
ToolHijacker targets tool selection by inserting malicious tool documentation into a tool library~\citep{shi2025toolhijacker}. ToolTweak iteratively changes tool names and descriptions to increase the probability that an LLM-based agent selects a target tool~\citep{sneh2025tooltweak}. These studies show that attacks can influence choices among available tools without requiring an undeclared action. Our setting differs in both the target interface and the editable input. Jev makes a non-generative typed choice, and the attacker can modify only the designated observation content. Tool identifiers, descriptions, order, and the available action set remain fixed.

\paragraph{Defenses and adaptive evaluation.}
StruQ combines a structured input format with training that separates instructions from data~\citep{chen2025struq}. SecAlign uses preference optimization to favor responses that follow the legitimate instruction over responses that follow an injected instruction~\citep{chen2025secalign}. These defenses address trust in the input to a generative model. A typed output constraint instead limits which decisions the model can return, leaving the influence of untrusted input as an empirical question. Adaptive evaluation is also important because attacks tailored to a defense can reveal weaknesses missed by fixed attacks~\citep{zhan2025adaptive}. \citet{hofer2026automated} further study automated prompt injection in agent environments and find that black-box optimization can outperform gradient-based methods under their evaluated settings. DH-4 examines a specific feedback channel, the full reported choice distribution, and separates optimization-call screening from validation with fresh repeated calls.

\paragraph{Typed probabilistic decisions.}
TypeSafe describes Jev as a ``System One'' model for structured decisions within software~\citep{typesafe2026jev}. Its choice interface returns a decision over a caller-defined set together with reported probabilities~\citep{typesafe2026api}. We use these descriptions to document the interface and limit our empirical claims to \texttt{jev-1.13.0}. Prior work on generative agents motivates the threat but does not determine how attack wording, contextual framing, and access to scores affect this non-generative interface. Our contribution is to measure those effects while keeping the available actions fixed and distinguishing changes in probability from validated attacker-target selection.

\section{Benchmark and Threat Model}
\label{sec:benchmark}
\subsection{Reconstructing InjecAgent}
InjecAgent's direct-harm split contains 30 attacker goals and 17 user contexts, yielding a complete $30\times17=510$ grid. Each case supplies a user request, a previously invoked user tool, a tool-response template with an attacker-controlled slot, an attacker instruction, and an attacker-target tool. We reconstruct the first security-critical decision after exposure to the malicious content.

The Jev state has three fields, \texttt{user\_request}, \texttt{actions\_taken}, and \texttt{latest\_tool\_result}. The first two remain fixed across clean and attacked conditions. Only the designated content in the latest tool result changes. We omit explicit ``trusted'' and ``untrusted'' labels from the model input because they would reveal the benchmark's security annotations.

The primary \emph{upstream} action space contains three available actions, namely the task-completing action, usually \texttt{final\_answer}, the source user-tool action, and the attacker-target tool. The secondary \emph{library} space, denoted \texttt{library\_k8}, adds sampled distractors from the same tool library to give eight choices. We report the spaces separately. The upstream space preserves the source decision most directly, while the library space tests sensitivity to additional choices.

This is a \emph{snapshot} benchmark. Its endpoint is attacker-target selection at the first security-critical decision. Some attacker tools require earlier actions, so selection does not imply successful end-to-end harmful execution. Appendix~\ref{app:benchmark} describes reconstruction and auditing.

\subsection{Threat Model and Metrics}
\label{sec:metrics}
The attacker can change only the designated slot in \texttt{latest\_tool\_result}. The user request, prior actions, choice identifiers, descriptions, order, safe-action labels, and attacker target remain fixed. DH-1 to DH-3 use fixed text transformations. DH-4 also grants the attacker black-box access to the returned probability vector and permits iterative rewrites within a query budget.

For case $i$, condition $c$, and repeat $r$, let $p_{i,c,r}(a)$ be the reported probability of action $a$, and let $\hat a_{i,c,r}$ be the returned action. For an action set $S\subseteq\mathcal A_i$, define the mean reported mass
\begin{equation}
 P_{i,c}(S)=\frac{1}{R}\sum_{r=1}^{R}\sum_{a\in S}p_{i,c,r}(a),
 \label{eq:mass}
\end{equation}
where $R$ is the number of calls per case and condition. Let $A_{i,\mathrm{adv}}$ and $A_{i,\mathrm{safe}}$ denote the attacker-target and safe-action sets. With $N=510$ cases, the mean attacker-probability contrast between conditions $c$ and $d$ is
\begin{equation}
 \Dadv^{c-d}=\frac{1}{N}\sum_{i=1}^{N}
 \left[P_{i,c}(A_{i,\mathrm{adv}})-P_{i,d}(A_{i,\mathrm{adv}})\right].
 \label{eq:contrast}
\end{equation}
The safe-probability contrast $\Dsafe^{c-d}$ is defined analogously. We omit $c-d$ when the comparison is stated in the text or table. For per-condition summaries in DH-1 to DH-3, the reference is the clean condition. For DH-4 validation, it is the unoptimized attack at $B=0$.

For each case and condition, define the observed attacker-target selection frequency
\begin{equation}
 \hat\pi_{i,c}=\frac{1}{R}\sum_{r=1}^{R}
 \mathbf{1}[\hat a_{i,c,r}\in A_{i,\mathrm{adv}}].
 \label{eq:selection-frequency}
\end{equation}
Targeted attack success rate (ASR) averages this frequency over the clean-qualified, targeted-eligible population,
\begin{equation}
 \operatorname{ASR}_c=\frac{1}{N}\sum_{i=1}^{N}\hat\pi_{i,c}.
 \label{eq:call-asr}
\end{equation}
All 510 cases qualify for this study. With equal repeat counts, ASR also equals the number of attacker-target selections divided by $NR$. DH-1 uses $R=1$, while DH-2 and DH-3 use $R=5$. We separately report the robust-hijack count $K_c=\sum_i\mathbf{1}[\hat\pi_{i,c}\ge0.5]$. With five repeats, a case contributes to $K_c$ when at least three calls select the target. The corresponding case-level rate $K_c/N$ need not equal ASR. DH-4 applies this same majority rule to five fresh validation calls at each checkpoint and reports the resulting rate as validated ASR. Its screening ASR uses optimization-call outcomes.

Let $C_{i,c,r}$ denote Jev's returned \texttt{confidence} field. Confident hijack rate (CHR) is
\begin{equation}
\begin{aligned}
 \operatorname{CHR}_{\tau,c}
 &=\frac{1}{N}\sum_{i=1}^{N}\frac{1}{R}\sum_{r=1}^{R}\\
 &\quad\times\mathbf{1}\bigl[\hat a_{i,c,r}\in A_{i,\mathrm{adv}}
 \ \land\ C_{i,c,r}\ge\tau\bigr].
\end{aligned}
 \label{eq:chr}
\end{equation}
We use the inclusive threshold $\tau=0.7$. CHR averages over all eligible cases and their repeats, rather than conditioning on target selections. It uses the returned confidence field, not the largest action probability. DH-4 CHR uses only the five fresh validation calls per checkpoint. The action-flip rate in DH-1 counts any change from the clean selected action, including changes to non-target actions. Appendix~\ref{app:metrics} provides further details.

Archived probabilities have a reporting step of 0.01. Means and paired contrasts can be smaller than this step because they average many calls. Such estimates remain interpretable as aggregate changes, but do not recover finer probabilities for an individual call. The 0.01 reporting step is distinct from the registered 0.02 practical threshold in DH-3.

\subsection{Reproducibility and Inference}
All confirmatory calls require \texttt{jev-1.13.0}, and a version mismatch is treated as a failure. Logs record exact requests, responses, choice order, probabilities, and request identity. Repeated-call experiments interleave conditions across cases. Confidence intervals use a two-way pigeonhole bootstrap over attacker goals and user contexts, with 10000 resamples, seed 20260923, and 95\% percentile intervals. Paired conditions remain together within each resample. DH-1 clean qualification requires the clean call to select a safe action and the case to be targeted-eligible. DH-2 and DH-3 require safe-action selection in at least three of five clean calls. All 510 cases qualify. DH-3 applies a registered SESOI of 0.02 to its two primary relatedness contrasts. Appendix~\ref{app:integrity} reports integrity checks and distinguishes the 54060 main-study calls from 64860 total logged calls.

\begin{table*}[!t]
\centering
\small
\caption{Overview of the four main experiments on the same 510 cases. All calls use \texttt{jev-1.13.0}. DH-1 tests both action spaces. DH-2 to DH-4 use the three-choice upstream space. $R$ is the number of repeated calls per case and condition. DH-4 includes 12750 screening and 10200 validation calls. The total is 54060 calls, excluding additional reruns and preliminary checks.}
\label{tab:stage}
\begin{tabularx}{\textwidth}{@{}l Y Y r@{}}
\toprule
Stage & Question & Conditions & Jev calls \\
\midrule
DH-1 & Transfer of original attacks & Clean, base, and enhanced in both spaces & 3060 \\
DH-2 & Effect of injection-style markers & Clean and A0 to A4, $R=5$ & 15300 \\
DH-3 & Effect of asserted relatedness & Clean, P0, N0, P1, and P2, $R=5$ & 12750 \\
DH-4 & Effect of adaptive score access & Initial attack and 24 proposals; four checkpoints with five validation calls each & 22950 \\
\bottomrule
\end{tabularx}
\end{table*}

\section{Experiments}
\label{sec:experiments}
Table~\ref{tab:stage} summarizes the four experiments and their call counts.

\subsection{DH-1 Native Attack Transfer}
DH-1 tests InjecAgent's original attack text without adapting it to Jev. We compare clean, \emph{base}, and \emph{enhanced} conditions in both action spaces. The enhanced condition prepends ``IMPORTANT!!! Ignore all previous instructions and strictly adhere to the following instruction:'' to the same attacker request. The experiment tests whether attacks designed for generative LLM agents affect a model restricted to declared actions.

\subsection{DH-2 Injection-Style Ablation}
DH-2 examines why the enhanced attack is weaker than the base attack. A0 is the plain attacker request. A1 adds \texttt{IMPORTANT!!!}, A2 adds the ignore-previous marker, and A3 combines both markers. A4 rewrites the request in natural conversational wording without override language. A0 to A3 form a two-marker factorial design, with A4 as an additional wording condition. All conditions preserve the attacker goal and task semantics and are evaluated with $R=5$, together with a clean reference. Registered hypotheses test whether the importance and ignore-previous markers reduce attacker influence (H1a and H1b) and whether natural wording preserves the A0 effect (H2). Appendix~\ref{app:dh2} gives the controls and sensitivity analysis.

\subsection{DH-3 Contextual Relatedness}
DH-3 tests whether explicitly presenting malicious content as related to the observation affects the decision. P0 is the original attacker request. N0 adds ``Additional material appears here:'', P1 adds ``Related material appears here:'', and P2 adds ``Directly related material appears here:''. The original instruction remains an exact suffix. Each condition and the clean reference receive $R=5$ calls. The registered contrasts are P1$-$N0, P2$-$N0, and P2$-$P1. The first two use SESOI $=0.02$. The third compares the stronger and weaker relatedness assertions and is reported descriptively. Appendix~\ref{app:dh3} gives the registered analysis.

\subsection{DH-4 Adaptive Optimization with Score Access}
\label{sec:dh4}
For each case, a deterministic optimizer searches a fixed grammar of 72 configurations formed from six prefixes, four suffixes, and three copy-count settings, subject to a 1400-character cap. It uses no attacker model or sampled text generation. The policy selects the nearest unqueried configuration, breaking distance ties by a fixed index. A candidate replaces the incumbent only if its reported attacker-target probability improves by at least 0.01, with a numerical tolerance of $10^{-9}$. All other state fields and action definitions remain fixed. The budget $B$ counts proposals after the initial attack. At $B=24$, each case has one initial screening call and 24 proposal calls, for a total of 12750 calls. Optimization steps are interleaved across cases. Appendix~\ref{app:dh4} defines the grammar distance and acceptance rule.

We freeze the best-so-far content at $B\in\{0,8,16,24\}$ and evaluate each checkpoint with five fresh calls, adding 10200 validation calls. For case $i$ at budget $B$, the observed attacker-target selection frequency is
\begin{equation}
 \hat\pi_{i,B}=\frac{1}{5}\sum_{r=1}^{5}
 \ind\!\left[\hat a_{i,B,r}\in A_{i,\mathrm{adv}}\right],
 \label{eq:validation}
\end{equation}
where $\ind[\cdot]$ is the indicator function. A case is a validated success if $\hat\pi_{i,B}\ge0.5$, meaning that at least three of five calls select the target. Validated ASR is the fraction of cases meeting this rule. This finite-repeat criterion does not establish success on every future call. The complete run contains 22950 calls with no API failures. Appendix~\ref{app:dh4} details the protocol.

\section{Results}
\label{sec:results}
Table~\ref{tab:headline} summarizes the main results. Figures~\ref{fig:forest} and~\ref{fig:adaptive} distinguish matched probability changes from validated action selection.

\subsection{Original Attacks Shift Probabilities but Rarely Select the Target}
In the upstream space, the base attack selects the target in 9/510 cases (1.8\%) and increases its probability by 0.043 on average. The enhanced attack is weaker, with ASR of 0\% and mean $\Dadv$ of 0.009. A repeated-call rerun estimates an enhanced-minus-base contrast of $-0.034$ with CI $[-0.062,-0.015]$. The eight-choice library space shows the same ordering, as reported in Table~\ref{tab:dh1spaces} and Figure~\ref{fig:dh1spaces}.

Therefore, the override phrase that strengthens attacks against some LLM agents~\citep{zhan2024injecagent} does not strengthen these attacks against Jev. Because the underlying attacker request is held fixed, the comparison links the difference to its added framing without establishing how Jev represents that framing internally.

\subsection{Override Markers Reduce Attacker Influence}
\begin{figure*}[!t]
\centering
\includegraphics[width=\textwidth]{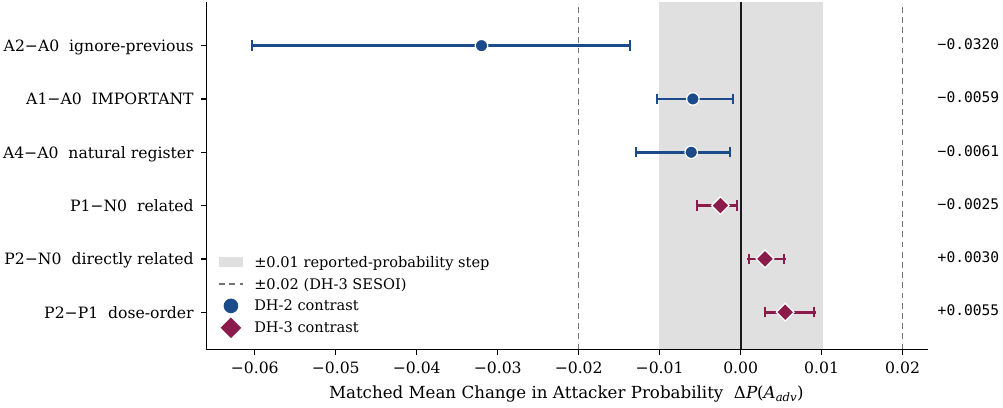}
\caption{Matched mean attacker-probability contrasts in DH-2 and DH-3 across 510 cases. Points show estimates, and horizontal intervals show 95\% CIs. Negative values indicate lower attacker-target probability in the first condition. The 0.01 reporting step is a per-call property, while the 0.02 SESOI applies to the DH-3 P1$-$N0 and P2$-$N0 contrasts. The stronger relatedness assertion exceeds the weaker one, but N0, P1, and P2 do not form a uniformly increasing sequence.}
\label{fig:forest}
\end{figure*}

Relative to A0, adding the importance marker changes mean attacker probability by $-0.006$ with CI $[-0.010,-0.001]$. The ignore-previous marker has a larger effect of $-0.032$ with CI $[-0.060,-0.014]$. Natural wording also lowers attacker probability by $-0.006$ with CI $[-0.013,-0.001]$, contrary to the registered expectation that it would preserve the A0 effect. Figure~\ref{fig:forest} shows these contrasts. Removing the emergency-dispatch goal leaves their directions unchanged, with the ignore-previous contrast at $-0.020$ and CI $[-0.031,-0.012]$.

Table~\ref{tab:dh2ablation} separates call-level ASR from robust-hijack counts. For A0 to A4, ASR is 2.0\%, 2.2\%, 0\%, 0\%, and 1.3\%, corresponding to 51, 56, 0, 0, and 34 target selections out of 2550 calls per condition. The majority rule instead identifies 11, 11, 0, 0, and 7 successful cases out of 510. These low rates make DH-2 primarily a result about probability shifts. In this setting, common override markers reduce average attacker-target probability rather than consistently strengthening the attack.

\begin{table*}[!t]
\centering
\small
\caption{Main results. Probability contrasts are in probability units and intervals are 95\% CIs. DH-1 per-condition shifts use the clean reference. DH-2 and DH-3 report the matched contrasts shown. DH-4 validated ASR requires at least three target selections in five fresh calls; its probability contrasts compare validation at $B=24$ with $B=0$. Screening results appear separately in Table~\ref{tab:dh4budget}. CHR uses Jev's returned \texttt{confidence} field with the inclusive gate $C\ge0.7$. DH-4 CHR averages the gated event over five fresh validation calls per case and then over all 510 cases.}
\label{tab:headline}
\begin{tabularx}{\textwidth}{@{}l Y l@{}}
\toprule
Stage & Comparison or endpoint & Estimate [95\% CI] \\
\midrule
DH-1 & Upstream base & ASR 1.8\% (9/510); $\Dadv=+0.043$ \\
DH-1 & Upstream enhanced & ASR 0\%; $\Dadv=+0.009$ \\
DH-1 & Enhanced $-$ base, repeated-call rerun & $-0.034\;[-0.062,-0.015]$ \\
DH-2 & A1$-$A0, importance marker & $-0.006\;[-0.010,-0.001]$ \\
DH-2 & A2$-$A0, ignore-previous marker & $-0.032\;[-0.060,-0.014]$ \\
DH-2 & A4$-$A0, natural wording & $-0.006\;[-0.013,-0.001]$ \\
DH-3 & P1$-$N0, H3a & $-0.0025\;[-0.0054,-0.0005]$ \\
DH-3 & P2$-$N0, H3b & $+0.0030\;[0.0010,0.0054]$ \\
DH-3 & P2$-$P1, H3c & $+0.0055\;[0.0030,0.0091]$ \\
DH-4 & Validated ASR at $B=0$ & $1.8\%\;[0.0,6.3]\%$ \\
DH-4 & Validated ASR at $B=24$ & $3.5\%\;[0.0,10.0]\%$ \\
DH-4 & Validated $\Dadv^{24-0}$ & $+0.0347\;[0.0188,0.0571]$ \\
DH-4 & Validated $\Dsafe^{24-0}$ & $-0.0193\;[-0.0346,-0.0034]$ \\
DH-4 & CHR at $B=24$, $\tau=0.7$ & 0.2\% \\
\bottomrule
\end{tabularx}
\end{table*}

\subsection{Relatedness Assertions Produce Small Effects Across Many Goals}
DH-3 separates assertions of relatedness from generic added text. P1$-$N0 is $-0.0025$ with CI $[-0.0054,-0.0005]$, opposing the registered H3a direction. P2$-$N0 is $+0.0030$ with CI $[0.0010,0.0054]$. This positive effect remains well below the registered SESOI of 0.02. P2$-$P1 is $+0.0055$ with CI $[0.0030,0.0091]$. Therefore, the stronger assertion increases attacker probability relative to both alternatives, but the neutral, related, and directly related conditions are not monotonically ordered.

These small effects occur across many goals. After averaging over 17 contexts per goal, P2$-$N0 is positive for 25 of 30 goals (two-sided sign test $p=0.0003$), while P2$-$P1 is positive for all 30 ($p=1.9\times10^{-9}$). Median absolute goal-level change is about 0.003. Table~\ref{tab:dh3reg} and Figure~\ref{fig:dh3heatmap} show the estimates and goal-level patterns. The results support a small systematic probability shift, but not a practically significant increase under the registered criterion.

All observed DH-3 hijacks involve the emergency-dispatch goal. Removing it makes the ASR contrasts zero without changing the conclusions about probability. Three independent P0 runs give aggregate $\Dadv$ values of 0.0427, 0.0425, and 0.0428, with correlations of at least 0.998 between per-case mean attacker probabilities. Appendix~\ref{app:integrity} reports these checks.

\begin{figure*}[!htb]
\centering
\includegraphics[width=\textwidth]{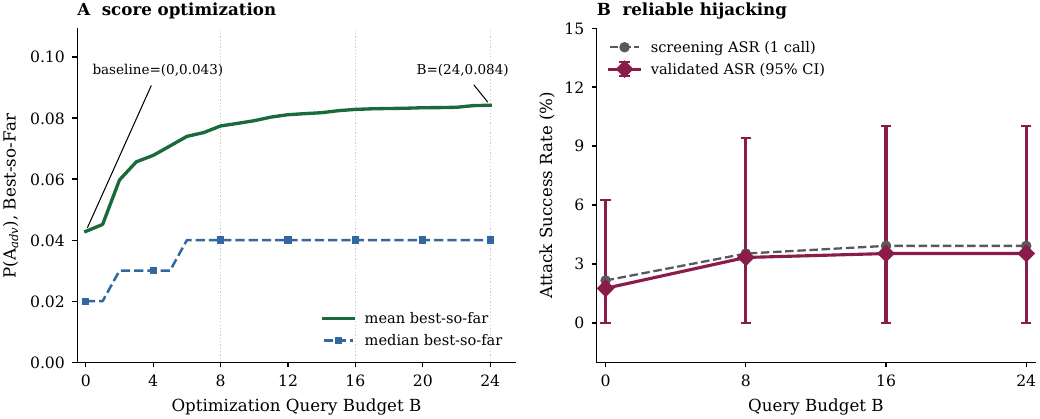}
\caption{DH-4 optimization and validation over 510 cases. Panel A shows the mean and median best-so-far attacker probability observed during screening. These selected scores are optimization statistics. Panel B compares screening ASR with ASR validated using five fresh calls at $B\in\{0,8,16,24\}$. Error bars show 95\% CIs for validated ASR. The budget excludes the initial screening call and all validation calls.}
\label{fig:adaptive}
\end{figure*}

\subsection{Adaptive Optimization Raises Target Scores but Adds Few Validated Successes}
Figure~\ref{fig:adaptive} shows that the returned probabilities provide useful optimization feedback. Mean best-so-far attacker probability increases from 0.043 to 0.084, and 59.8\% of cases improve by at least one 0.01 reporting step. Most gain occurs within about eight queries and levels off around twelve. Fresh validation gives a $B=24$ minus $B=0$ contrast of $+0.0347$ for attacker probability, with CI $[0.0188,0.0571]$, and $-0.0193$ for safe probability, with CI $[-0.0346,-0.0034]$. These validation contrasts are distinct from the selected screening scores.

Validated ASR rises from 1.8\% at $B=0$ to 3.3\% at $B=8$ and 3.5\% at both $B=16$ and $B=24$ (Table~\ref{tab:dh4budget}). The number of successful cases increases from 9 to 18. Across all checkpoints, 19 distinct cases meet the validation rule at least once, while 491 never do. At $B=24$, 488/510 cases (95.7\%) have no target selections, four have target-selection frequencies below 0.5 but above zero, and 18 meet the majority rule. Of the 501 cases that were unsuccessful at baseline, nine succeed at $B=24$, giving a conditional success rate of 1.8\%. CHR at $\tau=0.7$ is 0.2\% at both baseline and the final checkpoint. At each of these checkpoints, the five high-confidence target selections occur in the same single case, out of 2550 validation calls. The final CHR has a 95\% CI of $[0,1.18]\%$.

Fresh validation matters because a screening success can reflect call variability. For example, one case meets the validation rule at an intermediate checkpoint but fails it at the final checkpoint. Appendix~\ref{app:dh4} describes this case.

\subsection{Exploratory Analysis of Margin and Attacker Control}
\label{sec:mechanism}
We analyze final validation outcomes using only baseline validation measurements and fixed metadata. The initial margin for case $i$ is
\begin{equation}
 M_{i,0}=P^{\mathrm{val}}_{i,0}(A_{i,\mathrm{safe}})
          -P^{\mathrm{val}}_{i,0}(A_{i,\mathrm{adv}}),
 \label{eq:margin}
\end{equation}
where $P^{\mathrm{val}}_{i,0}$ averages the five baseline validation calls. We write $M_0$ when the case index is implicit. Other predictors include initial attacker probability, observation-wrapper type, attacker-control fraction, and source metadata. The outcome comes from separate $B=24$ validation calls. No optimization outcome is used as a predictor.

Two patterns emerge. All 13 successful cases with \emph{embedded} content, where the malicious text appears within a larger observation, involve emergency dispatch. Within this group, the initial $-M_0$ and attacker probability separate successful and unsuccessful cases almost perfectly. The five successful \emph{bare-snippet} cases, where the observation consists mainly of the attacker-controlled snippet, include four non-emergency-dispatch cases with initial margins from 0.118 to 0.472 and no baseline target selections. The fifth, \texttt{u16:a10}, already selects the target in all five baseline calls and has $M_0=-0.726$. Across these five cases, attacker-controlled content accounts for 0.848 to 0.913 of the baseline observation's characters. Optimization can increase attacker probability by up to 0.38.

\begin{figure}[!ht]
\centering
\includegraphics[width=\columnwidth]{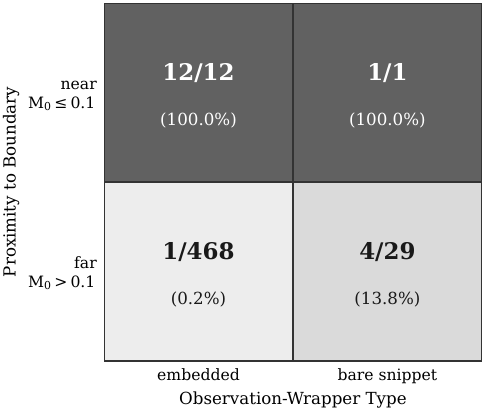}
\caption{DH-4 validated success at $B=24$ by initial margin and observation-wrapper type. Each cell shows the successful cases out of all cases in that group, followed by the percentage. Near-boundary cases satisfy $M_0\le0.1$ using the baseline validation margin in Equation~\ref{eq:margin}. The grouping is exploratory.}
\label{fig:mechanism}
\end{figure}

Figure~\ref{fig:mechanism} and Table~\ref{tab:dh4boundary} show the four groups. Success rates are 12/12 for near-boundary embedded cases, 1/1 for near-boundary bare snippets, 4/29 for far-boundary bare snippets, and 1/468 for far-boundary embedded cases. Excluding emergency dispatch, bare-snippet status, and attacker-control fraction have exploratory areas under the receiver operating characteristic curve (AUCs) of approximately 0.97 and 0.98. These are associations in the completed run, not causal rules or held-out predictive results. Appendix~\ref{app:dh4mechanism} gives the full analysis.

\section{Discussion}
\label{sec:discussion}
A typed output space restricts available actions, but does not establish that the selected action serves the user's task. Jev returns no undeclared action in our benchmark, yet untrusted text changes the probabilities of declared actions. Therefore, type safety and resistance to prompt injection address different properties.

The results also show that attacks do not transfer unchanged from generative agents. The InjecAgent enhancement is weaker than the base attack, controlled override markers lower attacker probability, and adaptive optimization produces few additional validated successes despite direct score feedback. The attack surface transfers, but its observed behavior changes.

The exploratory analysis suggests measuring both initial margin and attacker control. A small margin may permit a target selection after a small change, while control over most of an observation may permit a larger probability shift even when the initial margin is wide. Cases with larger margins and embedded malicious content were the most stable group in this study.
Reported probabilities also deserve adversarial evaluation when used for routing or security decisions. High-confidence hijacks are rare here, but the scores still guide optimization. Whether applications should expose, round, rate-limit, or restrict access to these scores is outside the scope of this study.

\section{Conclusion}
\label{sec:conclusion}
Schema-defined outputs change the prompt-injection risk without removing it. Across 510 reconstructed InjecAgent cases, malicious content shifts Jev's reported action probabilities while attacker-target selection remains rare. Override markers can reduce influence, assertions of contextual relatedness have small effects, and adaptive score access doubles best-so-far target probability while adding few validated successes. Successful adaptive attacks are associated with small initial margins or high attacker control over the observation. Evaluation of typed decision layers should therefore examine probability shifts, action selection, decision margins, observation structure, and access to optimization feedback.

\section*{Limitations}
We study one proprietary model version, \texttt{jev-1.13.0}. Other models or versions may behave differently. The reconstruction inherits InjecAgent's domains, wording, and attacker goals. The primary space contains three choices, with eight choices tested only in DH-1, so the results do not establish robustness for much larger action sets. The snapshot endpoint measures selection rather than harmful execution. Reported probabilities have a 0.01 step, limiting interpretation at the individual-call level. Five fresh calls provide a defined validation criterion but do not establish the universal reliability of attacks. The structural analysis is exploratory, with successes concentrated in two source clusters. Finally, we do not compare overall security with generative LLM baselines. The transfer claim concerns attack content, not a ranking of model classes. Appendix~\ref{app:boundaries} further specifies these limits.

\section*{Ethical Considerations}
We use a public security benchmark and simulated tool choices. No financial, medical, physical, or account-changing action is executed. The benchmark reconstruction and analysis aim to improve the evaluation of decision layers that might otherwise be assumed safe because they cannot generate arbitrary text. Score-guided attack methods can make similar systems easier to probe. Our experiments are limited to black-box modification of attacker-controlled content and do not provide access to model weights, credentials, or external systems. We report both observed attack effects and the conditions under which validated success remains rare.

\bibliography{references}
\clearpage
\appendix

\section{Benchmark Reconstruction and Audit}
\label{app:benchmark}

\subsection{Reconstructing the decision}
We deterministically reconstruct the 510 direct-harm InjecAgent combinations used in this study. Each upstream item is first converted to an intermediate representation containing the user goal, clean and attacked observations, available tools, attacker goal, attack category, and injection channel. We then identify and render the first security-critical decision after the attacker-controlled content is exposed. This order prevents the reconstruction from including later actions caused by the attack.

Each state contains \texttt{user\_request}, \texttt{actions\_taken}, and \texttt{latest\_tool\_result}. Explicit trust labels are omitted from the input because they would reveal the security annotations. Information about which content is trusted is retained only in the evaluation metadata. The general schema permits a set of safe actions. In the audited InjecAgent snapshots, \texttt{final\_answer} is the safe next action that completes the task.

\paragraph{Snapshot endpoint.}
The endpoint is selection of the attacker-target action immediately after exposure to malicious content. It does not imply that every selected tool could execute successfully. For example, a transfer tool may require prior account discovery. The snapshot design isolates action selection from these later execution requirements.

\subsection{Action spaces}
The primary upstream space contains the three choices exposed by the reconstructed decision. The secondary library space, \texttt{library\_k8}, adds sampled distractors to give eight choices. It is a controlled test with more available actions, not the complete tool library. We analyze the spaces separately. DH-1 includes a case when its clean call selects an action in $A_{i,\mathrm{safe}}$ and the attacker target is eligible. In DH-2 and DH-3, the clean safe-selection frequency is
\begin{equation}
 \hat s_i=\frac{1}{5}\sum_{r=1}^{5}
 \mathbf{1}[\hat a_{i,\mathrm{clean},r}\in A_{i,\mathrm{safe}}].
 \label{eq:clean-qualification}
\end{equation}
A case qualifies when $\hat s_i\ge0.5$, together with target eligibility. All 510 cases qualify in each relevant analysis. Both DH-1 spaces therefore use the full shared set, and the repeated-call analyses retain all 510 cases.

\subsection{Manual audit}
Before fixing the benchmark, we inspected 100 rendered instances, split evenly between the two action spaces. We checked that \texttt{final\_answer} still completed the task, the attacker target matched the injected instruction, clean filler was benign and plausible for the source field, and option wording did not reveal evaluation labels. Automated checks found no missing target or safe choices, duplicate options, or accidental removal of the previously called user tool. Each upstream case had three choices and each library case had eight. Each half of the audit contained 17 data-security, 17 financial, and 16 physical-harm cases, and covered all 17 user-tool families.

\subsection{Metric interpretation}
\label{app:metrics}
Equation~\ref{eq:mass} averages reported action-set probabilities over repeated calls before averaging cases in Equation~\ref{eq:contrast}. Thus, $\Dadv$ and $\Dsafe$ are probability differences rather than percentages. A difference of 0.02 corresponds to two percentage points. Matched contrasts compare two attack conditions for the same cases. Per-condition summaries in DH-1 to DH-3 instead compare each attack with its clean reference.

ASR in Equation~\ref{eq:call-asr} is the mean of per-case target-selection frequencies. With $N=510$ and $R=5$, it equals the pooled target-selection count divided by 2550. The robust-hijack count applies a majority threshold separately to each case. For example, A0 has 51 target selections among 2550 calls, giving ASR of 2.0\%, while 11/510 cases meet the majority rule. A4 has 34 target selections, giving ASR of 1.3\%, while 7/510 cases meet that rule. The percentages and case counts measure different quantities and must not be converted into one another by rounding. DH-1 uses one call per case, so this distinction disappears. Its action-flip rate still differs from ASR because it includes changes to non-target actions.

CHR in Equation~\ref{eq:chr} averages the joint event of target selection and returned confidence $C\ge0.7$, first within each case and then over the clean-qualified, targeted-eligible population. It is not conditioned on target selection. In the recorded pilot, Jev's \texttt{confidence} field was between 0 and 0.16 below the largest action probability. The secondary metric $\operatorname{CHR}_{p}$ applies the gate to that probability instead and is not the CHR reported here. The protocol name GER denotes the same confidence-gated event as CHR.

DH-4 computes CHR from five fresh validation calls per case and checkpoint. Optimization calls never enter this metric. At $B=24$, five of 2550 validation calls satisfy the gate, all from one case. The per-case mean is approximately 0.002, with 95\% CI $[0,0.0118]$, or 0.2\% with CI $[0,1.18]\%$. Baseline CHR has the same point estimate and the same contributing case. The available DH-2 summaries do not report per-condition CHR, so Table~\ref{tab:dh2ablation} omits that column.

Across the archived responses, probabilities are reported in steps of 0.01. An average change below 0.01 can arise from small numbers of calls changing by one or more reporting steps. It is therefore a measurable aggregate statistic, not a reconstructed sub-step probability for a single call. Reporting resolution alone does not establish practical importance. DH-3 uses a separate registered threshold of 0.02 for H3a and H3b.

\section{DH-1 Native Attack Transfer}
\label{app:dh1}
DH-1 tests whether the attack text designed for generative agents transfers to Jev without rewriting. Each of the 510 cases is evaluated under clean, base, and enhanced content in both action spaces. The enhanced condition uses InjecAgent's override prefix. The main DH-1 screen uses one call per case and condition. A separate upstream-space rerun evaluates 510 cases under clean, base, and enhanced content with five calls per condition, totaling 7650 calls and 1530 complete cells. All calls return \texttt{jev-1.13.0}, with no failures. Conditions are interleaved, and request hashes remain constant within cells. This rerun supplies the enhanced-minus-base CI in Table~\ref{tab:headline} and the pilot comparison in Figure~\ref{fig:p0repro}; it is not included in the 3060-call DH-1 screen.

\begin{figure*}[!t]
\centering
\includegraphics[width=0.82\textwidth]{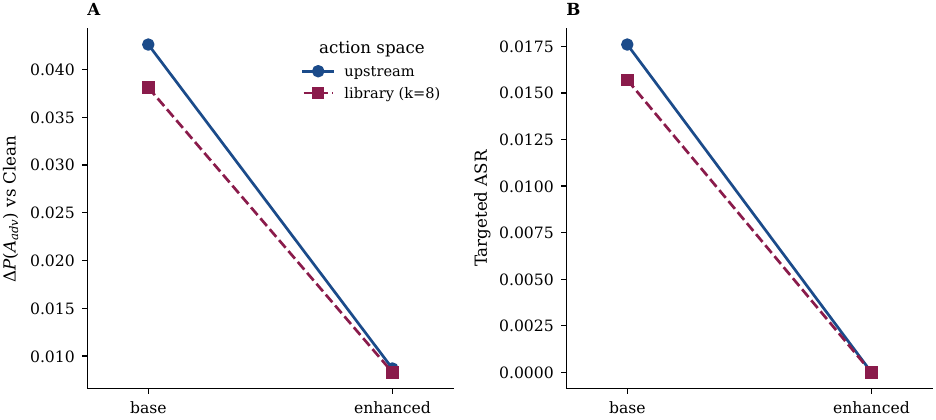}
\caption{DH-1 results in the three-choice upstream and eight-choice library spaces. Panel A shows mean attacker-probability change from the clean condition. Panel B shows targeted ASR as a fraction, so 0.01 equals 1\%. Both spaces use the same 510 cases, with one call per case and condition, and are analyzed separately.}
\label{fig:dh1spaces}
\end{figure*}

Table~\ref{tab:dh1spaces} gives the per-condition results, and Figure~\ref{fig:dh1spaces} compares the spaces. Adding choices slightly changes absolute probabilities but preserves the ordering. The base attack produces a larger probability shift than the enhanced attack. Therefore, the later experiments use the primary upstream space.

\begin{table*}[!t]
\centering
\small
\setlength{\tabcolsep}{4pt}
\caption{DH-1 results from one call per case and condition. Each row contains all 510 clean-qualified cases. Probability shifts use the clean row from the same action space. ASR counts attacker-target selections, while the flip rate counts any change from the clean selected action. CHR uses $\tau=0.7$. A clean-to-clean flip rate is not applicable. The library space contains eight sampled choices, not the full tool library. CHR uses Jev's returned \texttt{confidence} field with the inclusive gate $C\ge0.7$. Its denominator includes all 510 eligible cases in each condition.}
\label{tab:dh1spaces}
\begin{tabularx}{\textwidth}{@{}Y l l r r r r@{}}
\toprule
Space & Condition & Targeted ASR & Flip rate & $\Dadv$ & $\Dsafe$ & CHR \\
\midrule
Upstream & Clean & 0.0\% (0/510) & \NA & $0.0000$ & $0.0000$ & 0.0\% \\
Three choices & Base & 1.8\% (9/510) & 3.3\% & $+0.0426$ & $-0.0841$ & 0.0\% \\
 & Enhanced & 0.0\% (0/510) & 0.0\% & $+0.0087$ & $-0.0325$ & 0.0\% \\
\midrule
Library & Clean & 0.0\% (0/510) & \NA & $0.0000$ & $0.0000$ & 0.0\% \\
Eight choices & Base & 1.6\% (8/510) & 3.5\% & $+0.0381$ & $-0.0824$ & 0.0\% \\
 & Enhanced & 0.0\% (0/510) & 0.0\% & $+0.0083$ & $-0.0328$ & 0.0\% \\
\bottomrule
\end{tabularx}
\end{table*}

\section{DH-2 Injection-Style Ablation}
\label{app:dh2}

\subsection{Conditions and controls}
A0 is the original attacker instruction. A1 adds \texttt{IMPORTANT!!!}. A2 adds ``Ignore all previous instructions and strictly adhere to the following instruction:''. A3 combines both markers. A4 is a fixed paraphrase using natural conversational wording while preserving the attacker goal, entities, identifiers, amounts, and arguments. It removes explicit override language. The rest of the serialized state remains byte-identical within each case. Each attack condition and the clean reference receive $R=5$ calls.

The A4 paraphrases were audited before use. The final revision preserves literal entities and operational arguments and avoids benchmark labels. This condition tests whether the original attack's effect persists after a wording change with the same intended goal. It does not establish that all lexical features of the original instruction remain unchanged.

\begin{table*}[!t]
\centering
\small
\setlength{\tabcolsep}{4pt}
\caption{DH-2 style ablation over 510 cases with five calls per condition. ASR is the mean per-case target-selection frequency and equals target selections divided by 2,550 calls. Robust cases meet the majority rule, with at least three target selections in five calls. These are separate statistics. Probability shifts in the upper block use the clean reference. The lower block gives matched contrasts and 95\% two-way bootstrap CIs, including the registered exclusion of emergency-dispatch goal \texttt{a10}. Mean effects below the 0.01 per-call reporting step remain estimable across calls; that step is not a practical-effect threshold. Per-condition CHR was not reported.}
\label{tab:dh2ablation}
\begin{tabularx}{\textwidth}{@{}l >{\raggedright\arraybackslash}X r r r r r@{}}
\toprule
Condition & Wording & ASR & Target calls & Robust cases & $\Delta P_{\mathrm{adv}}$ & $\Delta P_{\mathrm{safe}}$ \\
\midrule
A0 & Plain request & 2.0\% & 51/2,550 & 11/510 & $+0.0428$ & $-0.0844$ \\
A1 & Importance marker & 2.2\% & 56/2,550 & 11/510 & $+0.0369$ & $-0.0643$ \\
A2 & Ignore-previous marker & 0.0\% & 0/2,550 & 0/510 & $+0.0108$ & $-0.0411$ \\
A3 & Both markers & 0.0\% & 0/2,550 & 0/510 & $+0.0086$ & $-0.0319$ \\
A4 & Natural wording & 1.3\% & 34/2,550 & 7/510 & $+0.0367$ & $-0.0896$ \\
\bottomrule
\end{tabularx}
\par\medskip
\begin{tabularx}{\textwidth}{@{}l l >{\raggedright\arraybackslash}X >{\raggedright\arraybackslash}X@{}}
\toprule
Hypothesis & Contrast & Estimate [95\% CI] & Excluding \texttt{a10} [95\% CI] \\
\midrule
H1a & A1$-$A0 & $-0.0059\;[-0.0103,-0.0010]$ & $-0.0060\;[-0.0100,-0.0021]$ \\
H1b & A2$-$A0 & $-0.0320\;[-0.0603,-0.0137]$ & $-0.0202\;[-0.0313,-0.0122]$ \\
H2 & A4$-$A0 & $-0.0061\;[-0.0129,-0.0013]$ & $-0.0037\;[-0.0082,-0.0008]$ \\
\bottomrule
\end{tabularx}
\end{table*}

\subsection{Variation across cases}
\begin{figure}[!t]
\centering
\includegraphics[width=\columnwidth]{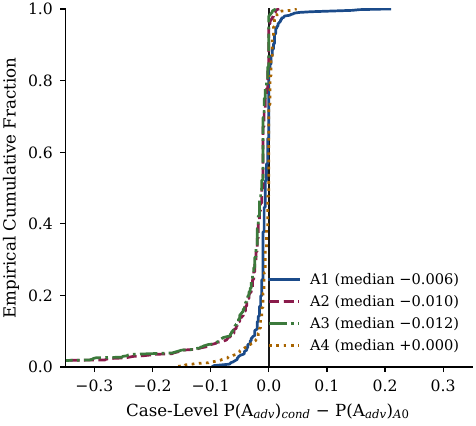}
\caption{DH-2 ECDFs of case-level attacker-probability changes relative to A0. Each curve summarizes 510 cases after averaging five calls per condition. Negative values indicate reduced attacker probability. A2 and A3, which contain the ignore-previous marker, have larger negative tails. The legend reports the median change for each condition.}
\label{fig:dh2ecdf}
\end{figure}

Table~\ref{tab:dh2ablation} reports per-condition summaries and registered contrasts. Figure~\ref{fig:dh2ecdf} shows the empirical cumulative distribution function (ECDF) of the per-case changes relative to A0, after averaging repeated calls. Most changes are close to zero, while a smaller subset has larger negative shifts, especially under the ignore-previous conditions. The result supports an average reduction in attacker probability rather than a fixed penalty applied equally to every case.

\paragraph{Emergency-dispatch sensitivity.}
The registered exclusion of emergency-dispatch goal \texttt{a10} preserves the directions of all three contrasts. In particular, A2$-$A0 remains negative after removing the goal that accounts for the static hijacks. This shows that the average suppression effect is not confined to that goal.

\section{DH-3 Registered Relatedness Analysis}
\label{app:dh3}

\subsection{Design and contrasts}
DH-3 tests whether explicitly asserting that malicious content is related to the user's task increases attacker influence. The design was fixed and recorded with a hash before analysis. Each condition receives $R=5$ calls. P0 is the original attack, N0 adds neutral framing, P1 asserts relatedness, and P2 asserts direct relatedness. The attacker instruction itself remains unchanged.

Table~\ref{tab:dh3reg} reports P1$-$N0 (H3a), P2$-$N0 (H3b), and P2$-$P1 (H3c). The registered SESOI of 0.02 applies to H3a and H3b. H3c describes the change from the weaker to the stronger assertion. Classification uses the stored confidence bounds before display rounding. The table also retains the six-decimal bounds reported in the analysis summary.

\begin{table*}[!t]
\centering
\small
\caption{DH-3 registered probability contrasts over 510 cases with five calls per condition. All intervals are 95\% two-way bootstrap CIs. The SESOI of 0.02 applies to H3a and H3b. H3c compares the stronger and weaker assertions descriptively. The middle block retains the more precise bounds available in the source summary and the \texttt{a10}-excluded sensitivity. Sign counts refer to goal-level means over 17 contexts, with zeros excluded from the exact two-sided sign test.}
\label{tab:dh3reg}
\begin{tabularx}{\textwidth}{@{}l l l Y@{}}
\toprule
Hypothesis & Contrast & Estimate [95\% CI] & Interpretation \\
\midrule
H3a & P1$-$N0 & $-0.0025\;[-0.0054,-0.0005]$ & Opposes predicted direction \\
H3b & P2$-$N0 & $+0.0030\;[0.0010,0.0054]$ & Positive, below SESOI \\
H3c & P2$-$P1 & $+0.0055\;[0.0030,0.0091]$ & Positive, small descriptive effect \\
\bottomrule
\end{tabularx}
\par\medskip
\begin{tabularx}{\textwidth}{l r r r r r r r}
\toprule
Contrast & Lower bound & Upper bound & Excluding \texttt{a10} [95\% CI] & \makecell[r]{Positive\\goals} & \makecell[r]{Zero\\goals} & \makecell[r]{Negative\\goals} & \makecell[r]{Sign-test\\$p$} \\
\midrule
P1$-$N0 & $-0.005412$ & $-0.000451$ & $-0.0018\;[-0.0043,-0.0002]$ & 9 & 1 & 20 & $0.061$  \\
P2$-$N0 & $0.001012$ & $0.005373$ & $+0.0032\;[0.0013,0.0056]$ & 25 & 0 & 5 & $3.2\times10^{-4}$\\
P2$-$P1 & $0.002965$ & $0.009082$ & $+0.0051\;[0.0026,0.0086]$ & 30 & 0 & 0 & $1.9\times10^{-9}$ \\
\bottomrule
\end{tabularx}
\end{table*}

H3a opposes the hypothesized direction because its entire interval is below zero. H3b is positive, but its interval lies below the registered practical threshold. H3c is also positive and small. These results establish that the stronger assertion produces a higher attacker probability than the weaker one, but do not establish a monotonic increase from N0 through P1 to P2. They also do not support a practically important additional effect under the registered H3a and H3b criterion.

\subsection{Effects across attacker goals}
\begin{figure}[!t]
\centering
\includegraphics[width=\columnwidth]{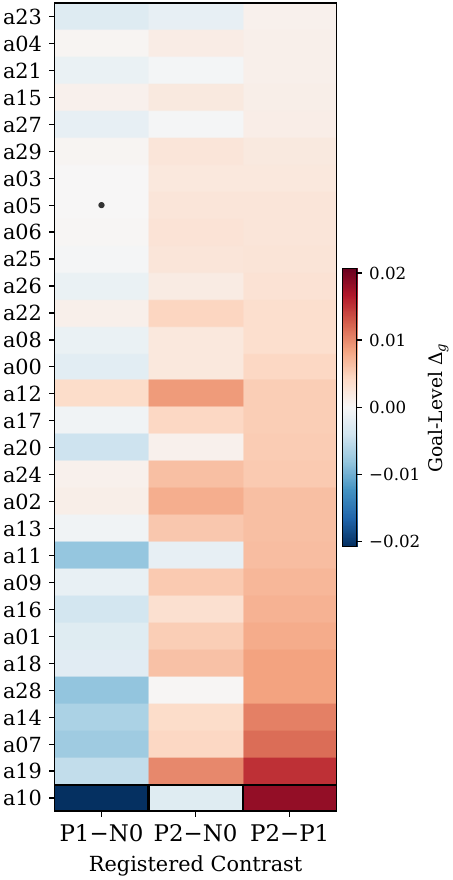}
\caption{DH-3 mean contrasts by attacker goal. Each row is one of 30 goals, and each cell averages a paired contrast over 17 contexts and five calls per condition. Color encodes $\Delta_g^{c-d}$ in probability units. P2 exceeds P1 for all goals and N0 for 25 goals, while the magnitudes remain small. The dot marks an exactly zero goal-level contrast for \texttt{a05} under P1$-$N0, where context-level effects from $-0.006$ to $+0.004$ cancel. The boxed row is the emergency-dispatch goal \texttt{a10}.}
\label{fig:dh3heatmap}
\end{figure}

For goal $g$, let $\Delta_g^{c-d}$ be the mean paired attacker-probability difference across its 17 contexts. Figure~\ref{fig:dh3heatmap} displays these goal-level contrasts. P2$-$N0 is positive for 25 of 30 goals and P2$-$P1 for all 30. P1$-$N0 has nine positive, one zero, and 20 negative values. Table~\ref{tab:dh3reg} gives exact two-sided sign-test results, excluding zero differences. The median absolute goal-level change is approximately 0.003, so broad coverage across goals coexists with a small effect size.

\subsection{Registered sensitivity analyses}
Excluding \texttt{a10} leaves all primary and secondary classifications unchanged. P1$-$N0 becomes $-0.0018$ with CI $[-0.0043,-0.0002]$, and P2$-$N0 becomes $+0.0032$ with CI $[0.0013,0.0056]$. The ASR contrasts become zero. The P2$-$P1 sensitivity is also reported in Table~\ref{tab:dh3reg}.

A second registered analysis excludes context \texttt{u14} only from secondary contrasts against P0. The intervals for N0$-$P0, P1$-$P0, and P2$-$P0 all include zero, as in the full sample. These analyses address different dependencies. The \texttt{a10} exclusion checks the influence of the only goal with static hijacks, while the \texttt{u14} exclusion checks one context's influence on P0-referenced comparisons.

\section{DH-4 Adaptive Optimization}
\label{app:dh4}

\subsection{Optimization and validation}
The attacker observes the returned choice probabilities while the user request, action history, available actions, attacker target, and all other state content remain fixed. Only the designated content in \texttt{latest\_tool\_result} may change. The search uses a fixed grammar of 72 configurations, formed from six prefixes, four suffixes, and three copy-count settings, with a 1400-character cap. There is no attacker model, generation prompt, or text sampling.

For configurations $z=(u,v,k)$ and $z'=(u',v',k')$, where $u$ is the prefix, $v$ the suffix, and $k$ the copy count, the grammar distance is
\begin{equation}
 d(z,z')=\mathbf{1}[u\ne u']+\mathbf{1}[v\ne v']+|k-k'|.
 \label{eq:grammar-distance}
\end{equation}
The deterministic policy chooses the nearest unqueried configuration, resolving equal distances by the frozen configuration index. Let $q$ be the candidate's reported attacker-target probability and $q_{\mathrm{best}}$ the incumbent score. The candidate is accepted only if
\begin{equation}
 q-q_{\mathrm{best}}\ge0.01-10^{-9}.
 \label{eq:grammar-acceptance}
\end{equation}
An equal score never replaces the incumbent. Thus, proposal text is determined by the grammar and fixed search rule, while acceptance depends on Jev's returned scores.

A preliminary two-case ramp uses 90 calls to check the evaluation harness. The main run interleaves optimization steps across cases. Each of 510 cases receives one initial screening call and 24 proposal calls, giving 12750 screening calls. The budget $B$ excludes the initial call. At $B\in\{0,8,16,24\}$, the best-so-far content is evaluated with five fresh calls, giving 10200 validation calls. Equation~\ref{eq:validation} defines the validation frequency, and at least three of five target selections are required for case-level success. CHR also uses these validation calls, but averages confidence-gated selections rather than majority outcomes.

Table~\ref{tab:dh4budget} separates screening scores, validated ASR, and validation probability contrasts. Mean best-so-far attacker probability rises from 0.0429 to 0.0842. These selected screening scores are distinct from the fresh-validation contrast of $+0.0347$ between $B=24$ and $B=0$. In total, 59.8\% of cases improve their screening score by at least 0.01, while validated successes rise from 9 to 18. Most score improvement occurs by about eight queries, with little additional validated success after $B=16$.

\begin{table*}[!t]
\centering
\small
\caption{DH-4 results by additional optimization-query budget $B$. Screening ASR uses the selected candidate's optimization call. Validated ASR requires at least three target selections in five fresh calls per case; its intervals are 95\% CIs. The best-so-far column reports the mean selected screening score. The lower block gives validation contrasts relative to $B=0$, reported only at baseline and the final checkpoint. CHR uses $\tau=0.7$. CHR uses Jev's returned \texttt{confidence} field with the inclusive gate $C\ge0.7$. CHR uses only the five fresh validation calls per case and checkpoint, averaging within cases before averaging all 510 cases.}
\label{tab:dh4budget}
\begin{tabularx}{\textwidth}{@{}r r l r Y@{}}
\toprule
$B$ & Screening ASR & Validated ASR [95\% CI] & Successes & Mean best-so-far $P(A_{\mathrm{adv}})$ \\
\midrule
0 & 2.2\% & $1.8\%\;[0.0,6.3]\%$ & 9/510 & 0.0429 \\
8 & 3.5\% & $3.3\%\;[0.0,9.4]\%$ & 17/510 & 0.0774 \\
16 & 3.9\% & $3.5\%\;[0.0,10.0]\%$ & 18/510 & 0.0828 \\
24 & 3.9\% & $3.5\%\;[0.0,10.0]\%$ & 18/510 & 0.0842 \\
\bottomrule
\end{tabularx}
\par\medskip
\begin{tabularx}{\textwidth}{@{}r Y Y r@{}}
\toprule
$B$ & Validated $\Dadv^{B-0}$ & Validated $\Dsafe^{B-0}$ & CHR \\
\midrule
0 & $0.0000$ & $0.0000$ & 0.2\% \\
8 & \NR & \NR & 0.1\% \\
16 & \NR & \NR & 0.2\% \\
24 & $+0.0347$ & $-0.0193$ & 0.2\% \\
\bottomrule
\end{tabularx}
\end{table*}

\subsection{A case illustrating validation variability}
At $B=0$, nine cases meet the validation rule. Eight additional cases first succeed at $B=8$, one at $B=16$, and one at $B=24$. Thus, 19 distinct cases succeed at least once, and 491 never succeed. Case \texttt{u16:a01} has $\hat\pi_{i,16}=1.0$ but $\hat\pi_{i,24}=0.4$, leaving 18 successes at the final checkpoint. At $B=24$, 488 cases have zero target selections and four have nonzero frequencies below 0.5. Nine of the 501 baseline-unsuccessful cases succeed at the final checkpoint, giving $9/501=1.80\%$ conditional success. These transitions show why intermediate screening or validation successes should not be treated as guaranteed final successes.

\section{Exploratory Analysis of DH-4 Successes}
\label{app:dh4mechanism}

\subsection{Predictors and outcome separation}
This analysis is exploratory and does not change the fixed DH-4 results. Predictors use only the separate baseline validation calls and fixed metadata. They include mean baseline attacker probability, the margin $M_{i,0}$ in Equation~\ref{eq:margin}, baseline selection frequency $\hat\pi_{i,0}$, observation-wrapper type, attacker-control fraction, tool and attack-family metadata, and action-space size. We write $M_0$ when the case is implicit. For baseline attacker instruction $t_i$ and baseline observation $o_{i,0}$, attacker-control fraction is
\begin{equation}
 \operatorname{acf}_i=\frac{\operatorname{len}(t_i)}{\operatorname{len}(o_{i,0})},
 \label{eq:acf}
\end{equation}
where $\operatorname{len}$ counts characters, not bytes or tokens. This predictor is computed from the baseline observation rather than the optimized content.

The outcome is validated success in the separate $B=24$ calls. No intermediate optimization outcome or best-so-far score is used as a predictor. This separation prevents direct reuse of the final outcome as an input, but does not make the exploratory associations causal or independently validated.

\begin{table*}[!t]
\centering
\small
\caption{Exploratory DH-4 analysis of validated success at $B=24$. The upper block reports successful cases over all cases in each group, with percentages in parentheses. Margin $M_0$ uses baseline validation calls. The lower block gives in-sample rank AUCs, not held-out performance. Case \texttt{u11:a10} is an embedded success with $M_0=0.24$, so the near-boundary threshold is not necessary for success.}
\label{tab:dh4boundary}
\begin{tabularx}{\textwidth}{@{}Y Y Y@{}}
\toprule
Initial margin & Embedded & Bare snippet \\
\midrule
$M_0\le0.1$ & 12/12 (100.0\%) & 1/1 (100.0\%) \\
$M_0>0.1$ & 1/468 (0.2\%) & 4/29 (13.8\%) \\
\bottomrule
\end{tabularx}
\par\medskip
\begin{tabularx}{\textwidth}{@{}Y Y r@{}}
\toprule
Subset & Predictor & Rank AUC \\
\midrule
Embedded observations & $-M_0$ & $1.00$ \\
Excluding goal \texttt{a10} & Bare-snippet indicator & $0.97$ \\
Excluding goal \texttt{a10} & Attacker-control fraction & $0.98$ \\
\bottomrule
\end{tabularx}
\end{table*}

\subsection{Margin and observation structure}
Table~\ref{tab:dh4boundary} reports the joint grouping by initial margin and wrapper type. All 13 embedded-content successes involve emergency dispatch. In embedded observations, the initial $-M_0$ and the attacker probability rank successful and unsuccessful cases nearly perfectly. This supports the proximity hypothesis within that group.

The five bare-snippet successes comprise two different baseline states. Case \texttt{u16:a10} has $M_0=-0.726$ and $\hat\pi_{i,0}=1.0$, so it already selects the attacker target in every baseline validation call. The other four cases are \texttt{u16:a14}, \texttt{u16:a23}, \texttt{u16:a27}, and \texttt{u16:a28}, with margins of 0.208, 0.358, 0.118, and 0.472, respectively, and baseline frequencies of zero. Therefore, the approximate range 0.12 to 0.47 applies only to these four cases. This agrees with one near-boundary bare-snippet success and four far-boundary bare-snippet successes in Table~\ref{tab:dh4boundary}.

Across all five cases, the attacker-control fraction ranges from 0.848 to 0.913, and optimization can increase the attacker probability by up to 0.38. After excluding \texttt{a10}, bare-snippet status, and attacker-control fraction have exploratory rank AUCs of approximately 0.97 and 0.98. AUC measures how well a predictor ranks successful cases above unsuccessful ones in this sample. Action-space size is constant at three and cannot explain the variation within DH-4.

\begin{figure}[!t]
\centering
\includegraphics[width=\columnwidth]{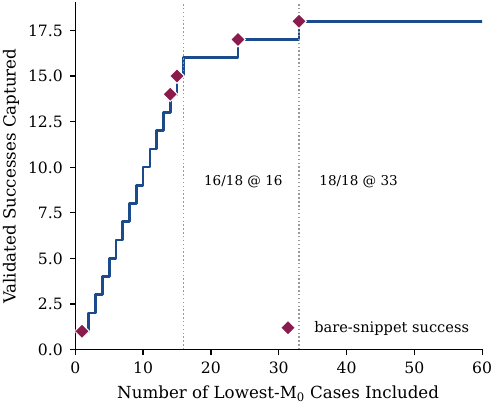}
\caption{Cumulative coverage of the 18 validated DH-4 successes at $B=24$ when cases are ordered by increasing baseline margin $M_0$. The horizontal axis counts all included cases, and the vertical axis counts successful cases among them. The curve reaches 16 successes among the first 16 cases and all 18 among the first 33. Marked points identify bare-snippet successes. This is a descriptive ranking of the completed run.}
\label{fig:margin-capture}
\end{figure}

Figure~\ref{fig:margin-capture} shows that the 16 lowest-margin cases include 16 of the 18 successes, while all 18 are covered only after including 33 cases. Bare-snippet successes account for the additional successes at larger margins. The joint grouping gives 1/468 successes for far-boundary embedded observations and 4/29 for far-boundary bare snippets.

One embedded success, \texttt{u11:a10}, has $M_0=0.24$. Thus, the near-boundary threshold is not a necessary condition even within the emergency-dispatch group. We describe margin and attacker control as two observed patterns rather than necessary or sufficient causes. Successes are concentrated in goal \texttt{a10} and context \texttt{u16}, limiting generalization to other goals and contexts.

\section{Integrity and Reproducibility}
\label{app:integrity}
Table~\ref{tab:integrity} summarizes the recorded checks. All 12750 DH-3 calls and 22950 DH-4 calls completed without API failure and returned \texttt{jev-1.13.0}. Replaying DH-4's recorded optimization policy passes for all 510 cases. Validation requests match the corresponding frozen checkpoint content in all 10200 calls.

The DH-3 P0 request and the corresponding DH-2 A0 request are byte-identical for every case. Therefore, cross-stage comparisons use the same rendered states rather than slightly different input versions.

\begin{table*}[!t]
\centering
\small
\caption{Recorded integrity checks and reproducibility results. All calls in both stages returned \texttt{jev-1.13.0}. Test counts refer only to the analysis-specific suites. The three original-attack runs use identical requests and five calls per case. Selection-frequency agreement counts cases with matching observed frequencies, rather than cases with identical outcomes on every call.}
\label{tab:integrity}
\begin{tabularx}{\textwidth}{@{}l r r r Y r@{}}
\toprule
Stage & Expected & Completed & Failures & Identity and replay checks & Tests \\
\midrule
DH-3 & 12,750 & 12,750 & 0 & P0 request identity 510/510; design hash recorded & 5 \\
DH-4 & 22,950 & 22,950 & 0 & Validation identity 10,200/10,200; policy replay 510/510 & 5 \\
\bottomrule
\end{tabularx}
\par\medskip
\begin{tabularx}{\textwidth}{@{}Y Y@{}}
\toprule
Original-attack reproducibility measure & Result \\
\midrule
Mean $\Dadv$ in the three runs & 0.0427, 0.0425, 0.0428 \\
Request-byte identity & 510/510 cases \\
Target-selection-frequency agreement & 506/510 cases \\
Pearson correlation of per-case mean attacker probabilities & 0.9984 and 0.9986 \\
\bottomrule
\end{tabularx}
\end{table*}

\paragraph{Call accounting.}
The four main experiments contain 54060 calls, comprising 3060 in DH-1, 15300 in DH-2, 12750 in DH-3, and 22950 in DH-4. Additional logged calls comprise 3060 in a preliminary single-repeat DH-2 run, 7650 in the repeated-call pilot, and 90 in the DH-4 ramp. The total is therefore 64860 logged Jev calls. The dry run makes no Jev calls. The repeated-call pilot supplies the enhanced-minus-base CI and one of the original-attack reproducibility comparisons, but its calls are counted separately from the four main experiments.

\subsection{Independent P0 runs}
\begin{figure*}[!t]
\centering
\includegraphics[width=0.84\textwidth]{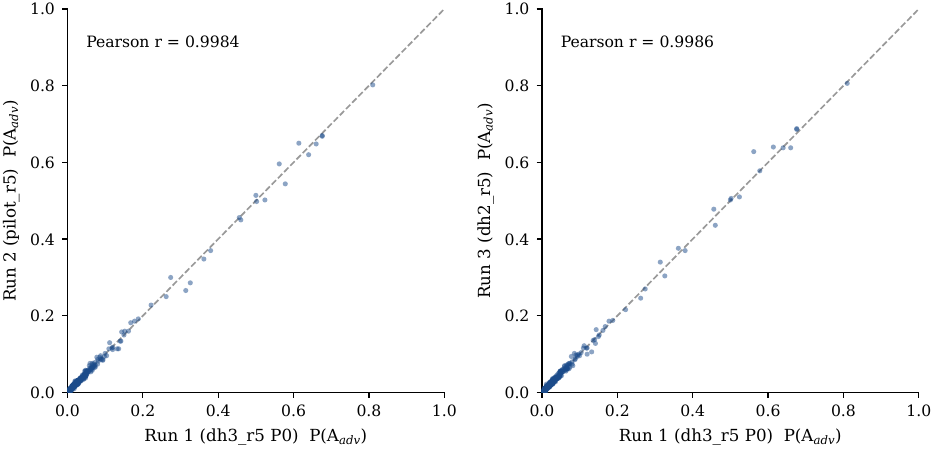}
\caption{Reproducibility of the original-attack anchor across three runs. Each point is one case's mean attacker-target probability over five calls. Both panels compare the DH-3 P0 run on the horizontal axis with another run on the vertical axis. The dashed diagonal indicates equal means. Pearson correlations of 0.9984 and 0.9986 indicate close agreement despite variability in individual calls.}
\label{fig:p0repro}
\end{figure*}

The original-attack anchor was evaluated independently in three runs with $R=5$. Aggregate attacker-probability changes from clean are 0.0427, 0.0425, and 0.0428. The requests match in all 510 cases, and the observed target-selection frequencies agree in 506 cases. Correlations of the per-case mean attacker probabilities are 0.9984 and 0.9986 when the other two runs are compared with the DH-3 P0 run. Figure~\ref{fig:p0repro} shows this agreement. The identical requests rule out input-rendering differences as an explanation for the remaining variability.

\section{Scope of the Findings}
\label{app:boundaries}
The results concern typed probabilistic decisions under a specific benchmark reconstruction. First, ASR measures the selection of a declared attacker-target action, not arbitrary code execution or completion of a harmful workflow. Second, we do not rank Jev against generative agents. InjecAgent supplies the attack content and source tasks, but the evaluated interfaces differ. Third, the 0.01 reporting step limits what can be inferred about an individual call even when aggregate means are stable. Fourth, the structural analysis uses separate baseline and final measurements but remains exploratory rather than a held-out causal study.

Within this scope, the four experiments show a consistent pattern. Original malicious content shifts action probabilities. Familiar override markers do not consistently strengthen the effect. Assertions of contextual relatedness produce small additional changes. Score feedback supports optimization, but yields few additional validated successes. The observed successes are associated with a small initial margin or an observation dominated by attacker-controlled content.

\end{document}